\documentclass[authoryear,12pt]{elsarticle}

\usepackage[margin=1in]{geometry}
\usepackage{setspace}
\usepackage{amsmath}
\usepackage{amsthm}
\usepackage{amssymb}
\usepackage{graphicx}
\usepackage{caption}
\usepackage{subcaption}
\usepackage{natbib}
\usepackage{lineno}

\begin{document}

\title{Selection for synchronized cell division in simple multicellular organisms}
\author{Jason~Olejarz$^\mathrm{a}$, Kamran~Kaveh$^\mathrm{a}$, Carl~Veller$^\mathrm{a,b}$, and Martin~A.~Nowak$^\mathrm{a,b,c}$}
\address{$^\mathrm{a}$Program for Evolutionary Dynamics, Harvard University, Cambridge, MA~02138 USA\\$^\mathrm{b}$Department of Organismic and Evolutionary Biology, Harvard University, Cambridge, MA~02138 USA\\$^\mathrm{c}$Department of Mathematics, Harvard University, Cambridge, MA~02138 USA\\jolejarz@fas.harvard.edu; kkavehmaryan@fas.harvard.edu; carlveller@fas.harvard.edu; martin\_nowak@harvard.edu}

\begin{abstract}
The evolution of multicellularity was a major transition in the history of life on earth. Conditions under which multicellularity is favored have been studied theoretically and experimentally. But since the construction of a multicellular organism requires multiple rounds of cell division, a natural question is whether these cell divisions should be synchronous or not. We study a simple population model in which there compete simple multicellular organisms that grow either by synchronous or asynchronous cell divisions. We demonstrate that natural selection can act differently on synchronous and asynchronous cell division, and we offer intuition for why these phenotypes are generally not neutral variants of each other.
\end{abstract}

\maketitle

\section*{Keywords}

Evolutionary Dynamics; Multicellularity; Synchronization; Cell Division

\section{Introduction}

The evolution of multicellular organisms from unicellular ancestors was one of the major transitions in the evolutionary history of life, and occurred at least 25 times independently, beginning as far back as 3--3.5 billion years ago \citep{Bonner_1988,Bonner_1998,Maynard_1995,Knoll_2003,Grosberg_2007,Rokas_2008,Niklas_2013}. Progress has been made in elucidating conditions that select for simple (undifferentiated) multicellularity over unicellularity, both in theory \citep{Pfeiffer_2003,Michod_2007,Niklas_2014,Olejarz_2014,Ghang_2014,Driscoll_2017,vanGestel_2017} and in experimental work \citep{Solari_2006a,Solari_2006b,Ratcliff_2012,Ratcliff_2013a,Tarnita_2015,Herron_2018}. 

Staying together and coming together are mechanisms for the formation of biological complexity \citep{Tarnita_2013}. Many simple multicellular organisms grow by the staying together (ST) of dividing cells, starting from a single progenitor cell \citep{Grosberg_1998}. Thus, a progenitor cell divides, and the daughter cell stays attached to the parent cell to form a complex of two cells (a `$2$-complex'). Further cell divisions produce new cells that stay with the group, and the organism grows. Eventually, when the organism is large enough, it starts to produce progenitor cells, which disperse to seed the growth of new multicellular organisms.

In simplest terms, the rate of growth (and ultimately the productivity) of a multicellular organism depends on the rate of division of its cells at various stages of its growth. For example, if a $k$-complex produces new cells $k$ times faster than a unicellular organism does, then the per-cell division rate of the multicellular organism is always equal to that of the unicellular organism, and multicellularity is no more productive than unicellularity  \citep{Bonner_1998,Willensdorfer_2009,Tarnita_2013}. But this represents only a single possibility. More generally, it is natural to consider cases where selection acts differently on complexes of different sizes \citep{Willensdorfer_2008,Willensdorfer_2009,Tarnita_2013}.  For example, if each $k$-complex produces new cells at a rate more than $k$ times faster than a unicellular organism, then the ST phenotype outcompetes the solitary phenotype, and multicellularity evolves. Natural selection may also act in non-linear, non-monotonic, or frequency-dependent ways on complexes of different sizes \citep{Celiker_2013,Julou_2013,Ratcliff_2013b,Lavrentovich_2013,Koschwanez_2013,Tarnita_2017}, and for many interesting cases, the population dynamics of ST is well characterized \citep{Michod_2005,Michod_2006,Allen_2013,Momeni_2013,Olejarz_2014,Ghang_2014,Maliet_2015,Kaveh_2016,vanGestel_2016}.

Against the background of this rich set of possibilities for the fitness effects of multicellularity, a question that has been ignored (to our knowledge) concerns the timing of cell divisions in the construction of a multicellular organism. Specifically, should their timing be independent or temporally correlated? That is, can there be selection for synchrony in cell division? Here, we study a model of simple multicellularity to determine the conditions under which synchronized cell division is favored or disfavored.

\section{Model}

We suppose that new cells remain attached to their parent cell after cell division.  This process continues until a complex reaches its maximum size, $n$.  A complex of size $n$ then produces new solitary cells. 

First, consider a population of asynchronously dividing cells.  For asynchronous cell division, the reproduction of each individual cell is a Poisson process, and cells divide independently.

For illustration, consider the case of neutrality.  The distribution of time intervals between production of new cells is exponential, with an average rate of a single cell division in one time unit.  In one time unit, on average, a single cell reproduces to form a complex containing two cells (the parent and the offspring).  With asynchronous cell division, it takes only another $1/2$ time unit, on average, for either of the cells of the $2$-complex to reproduce and form a $3$-complex.  Once the $3$-complex appears, in another $1/3$ time unit, on average, one of the three cells of the $3$-complex will reproduce to form a $4$-complex.  If $n=4$, then each $4$-complex produces new solitary cells at a rate of $4$ cells per time unit, and the cell division and staying together process starting from each new solitary cell is repeated.

Next, consider a population of synchronously dividing cells.  For synchronous cell division, all cells in a $k$-complex divide simultaneously, and simultaneous division of an entire cluster of cells is a Poisson process.

The growth process starting from a single cell is subtlely different if cells divide synchronously.  For illustration, again consider the case of neutrality.  The distribution of time intervals between doubling of an entire cluster of cells is exponential, with an average rate of one doubling of a cluster's size in a single time unit.  In one time unit, on average, a single cell reproduces to form a $2$-complex.  In one time unit, on average, the two cells in the $2$-complex simultaneously divide, the result being a new complex with four cells---the two parent cells and the two offspring.  (Notice that $3$-complexes do not form if cell division is perfectly synchronous.)  If $n=4$, then each $4$-complex produces new solitary cells at a rate of $4$ cells per time unit, and each new solitary cell repeats the cell division and staying together process.

\section{Results}

\subsection{$n=4$ cells}

We begin by studying the evolutionary dynamics for $n=4$.  The dynamics of asynchronous cell division and staying together for $n=4$ are described by the following system of differential equations:
\begin{equation}
\begin{pmatrix} \dot{y}_1 \\ \dot{y}_2 \\ \dot{y}_3 \\ \dot{y}_4 \end{pmatrix} = \begin{pmatrix} -\alpha_1 & 0 & 0 & 4\alpha_4 \\ \alpha_1 & -2\alpha_2 & 0 & 0 \\ 0 & 2\alpha_2 & -3\alpha_3 & 0 \\ 0 & 0 & 3\alpha_3 & 0 \end{pmatrix} \begin{pmatrix} y_1 \\ y_2 \\ y_3 \\ y_4 \end{pmatrix} - \phi_y(\vec{\alpha};\vec{y}) \begin{pmatrix} y_1 \\ y_2 \\ y_3 \\ y_4 \end{pmatrix}.
\label{eqn:n=4_asynch}
\end{equation}
The notation $\dot{y}_i$ indicates the time derivative.  Here, the variables $y_1$, $y_2$, $y_3$, and $y_4$ denote the frequencies of complexes with $1$, $2$, $3$, and $4$ asynchronously dividing cells, respectively.  The parameters $\alpha_k$ for $1 \leq k \leq 4$ represent the consequences of staying together on the fitness of cells in $k$-complexes.  We use the shorthand notation $\vec{\alpha}=\{\alpha_k\}$ to denote the vector of $\alpha_k$ values.  We choose $\phi_y(\vec{\alpha};\vec{y})$ such that $y_1+y_2+y_3+y_4=1$ at all times.  We obtain
\begin{equation}
\phi_y(\vec{\alpha};\vec{y}) = 1 + \sum_{k=1}^4 ky_k(\alpha_k-1).
\label{eqn:phi_A_4}
\end{equation}

The dynamics of synchronous cell division and staying together for $n=4$ are described by the following system of differential equations:
\begin{equation}
\begin{pmatrix} \dot{x}_1 \\ \dot{x}_2 \\ \dot{x}_4 \end{pmatrix} = \begin{pmatrix} -\alpha_1 & 0 & 4\alpha_4 \\ \alpha_1 & -\alpha_2 & 0 \\ 0 & \alpha_2 & 0 \end{pmatrix} \begin{pmatrix} x_1 \\ x_2 \\ x_4 \end{pmatrix} - \phi_x(\vec{\alpha};\vec{x}) \begin{pmatrix} x_1 \\ x_2 \\ x_4 \end{pmatrix}.
\label{eqn:n=4_synch}
\end{equation}
Here, the variables $x_1$, $x_2$, and $x_4$ denote the frequencies of complexes with $1$, $2$, and $4$ synchronously dividing cells, respectively.  (The parameters $\alpha_k$ for $1 \leq k \leq 4$ do not depend on synchronization or asynchronization in cell division.  Therefore, $\vec{\alpha}=\{\alpha_k\}$ is defined exactly as for the case of asynchronous cell division, as described above, although in the case of synchronization, the $\alpha_3$ value in $\vec{\alpha}$ is irrelevant.)  We choose $\phi_x(\vec{\alpha};\vec{x})$ such that $x_1+x_2+x_4=1$ at all times.  We obtain
\begin{equation}
\phi_x(\vec{\alpha};\vec{x}) = 1 + \sum_{p=0}^2 kx_k\left(\alpha_k-1\right), \mathrm{\; where \;} k=2^p.
\label{eqn:phi_S_4}
\end{equation}

In what follows, we use an asterisk to denote quantities that are in steady-state.  For asynchronously dividing cells, $\{y_k^*\}$ and $\phi_y^*(\vec{\alpha})$ denote the frequencies of $k$-complexes and the population fitness, respectively, when $\dot{y}_k=0$ for all $k$.  For synchronously dividing cells, $\{x_k^*\}$ and $\phi_x^*(\vec{\alpha})$ denote the frequencies of $k$-complexes and the population fitness, respectively, when $\dot{x}_k=0$ for all $k$.

\begin{figure}
\centering
\includegraphics*[width=0.9\textwidth]{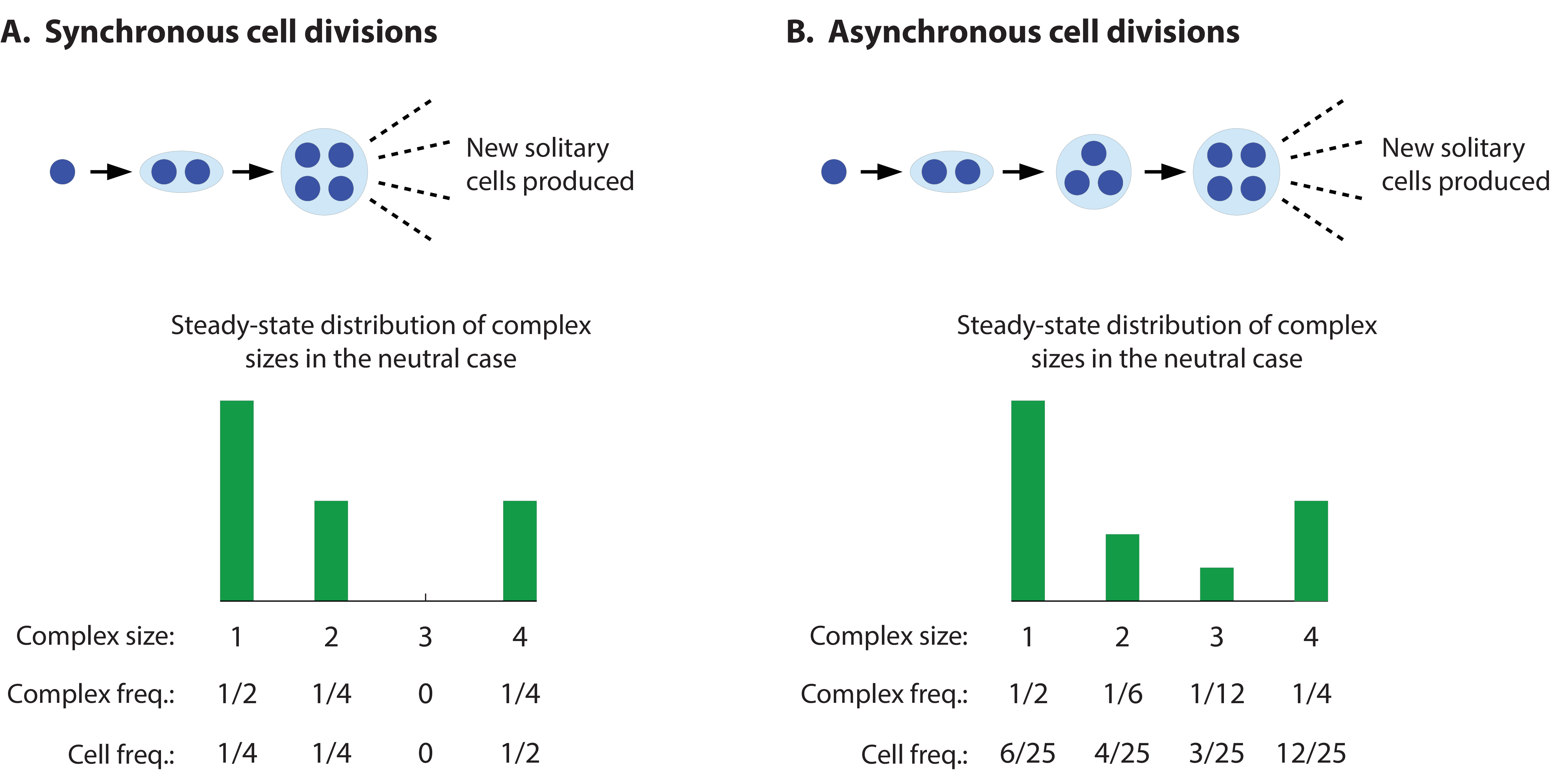}
\caption{Growth of multicellular organisms by synchronous and asynchronous cell division, when maximum size is $n=4$ cells.  (A) In synchronous cell division, a solitary cell divides to form a $2$-complex, and then both cells of the $2$ complex divide simultaneously to form a $4$-complex. Further cell divisions in the $4$-complex give rise to dispersing single cells. When all cells are equally productive in terms of their own division rates, no matter what size complex they are in (the neutral case), a steady state is reached where $1/2$ of organisms are single cells (accounting for $1/4$ of all cells), $1/4$ are $2$-complexes ($1/4$ of all cells), and $1/4$ are $4$-complexes ($1/2$ of all cells). $3$-complexes are never produced. (B) In asynchronous cell division, a solitary cell divides to form a $2$-complex, after which one of the cells in the $2$-complex divides to form a $3$-complex, after which one of the cells in the $3$-complex divides to form a $4$-complex. Further divisions lead to dispersing single cells. In the neutral case, an asynchronous population reaches a steady state where $1/2$ of complexes are single cells ($6/25$ of all cells), $1/6$ are $2$-complexes ($4/25$ of all cells), $1/12$ are $3$-complexes ($3/25$ of all cells), and $1/4$ are $4$-complexes ($12/25$ of all cells). }
\label{fig:n=4}
\end{figure}

The processes of staying together with synchronous and asynchronous cell division for the case $n=4$ are shown schematically in Figure \ref{fig:n=4}.

Suppose that we have a mixed population of cells with the synchronously and asynchronously dividing phenotypes.  Notice that there is competitive exclusion.  If $\phi_x^*<\phi_y^*$, then asynchronously dividing cells outcompete synchronously dividing cells.  If $\phi_x^*>\phi_y^*$, then synchronously dividing cells outcompete asynchronously dividing cells.  If $\phi_x^*=\phi_y^*$, then synchronously and asynchronously dividing cells can coexist.

From Equations \eqref{eqn:phi_A_4} and \eqref{eqn:phi_S_4}, for the particular case $\alpha_1=\alpha_2=\alpha_3=\alpha_4=1$, we have $\phi_x(1,1,1,1;\vec{x})=\phi_y(1,1,1,1;\vec{y})=\phi_x^*(1,1,1,1)=\phi_y^*(1,1,1,1)=1$, and there is neutrality between the synchronously and asynchronously dividing phenotypes.  But the case $\alpha_1=\alpha_2=\alpha_3=\alpha_4=1$ is nongeneric.  What happens if $\alpha_k \neq 1$ for some $k$?

\begin{figure}
\centering
\includegraphics*[width=0.7\textwidth]{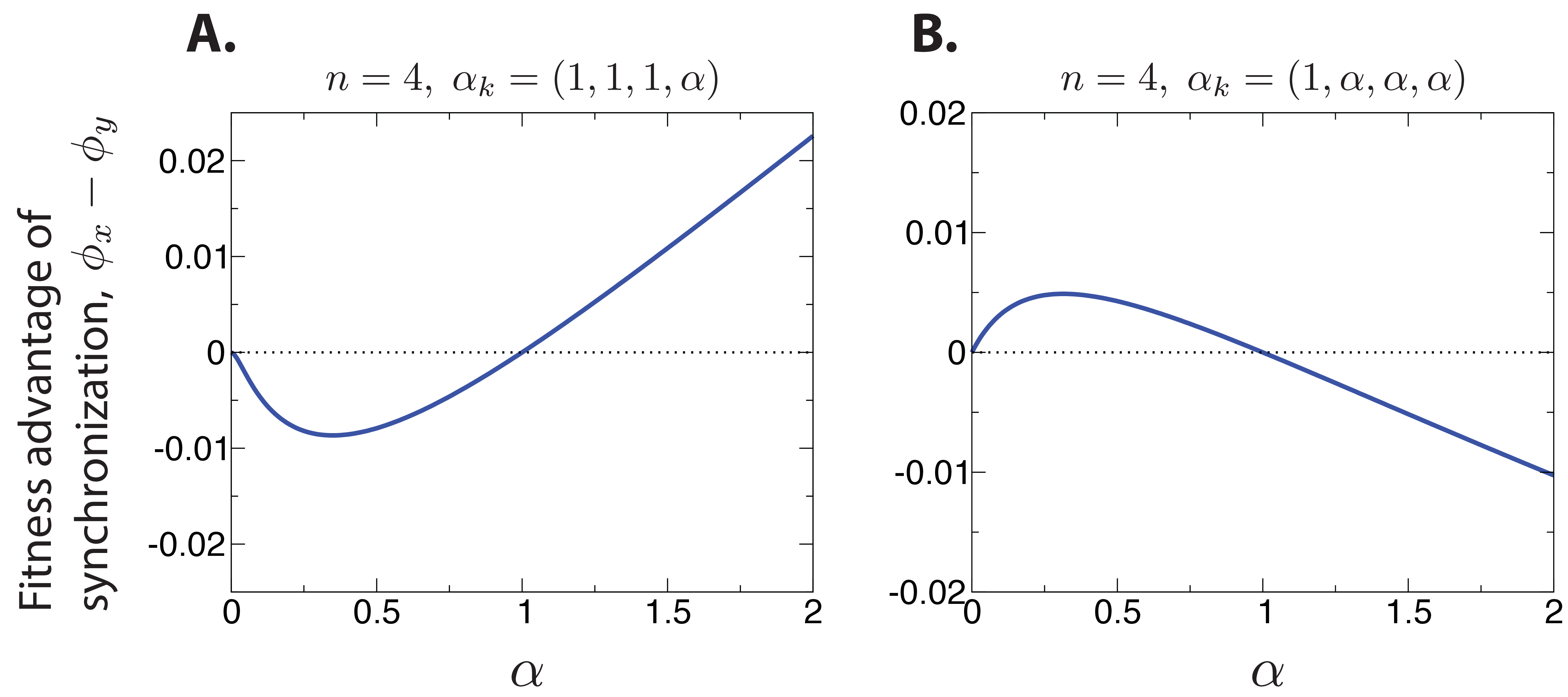}
\caption{The differences in the steady-state fitnesses of the synchronously and asynchronously dividing populations are shown for two choices of $\vec{\alpha}$ for $n=4$.  When multicellularity is selected for ($\alpha > 1$), synchronization of cell divisions is the more successful phenotype in some cases (A), and asynchronization in others (B).}
\label{fig:alpha_4}
\end{figure}

To understand the effect of $\alpha_k$ on the evolutionary dynamics, we consider a couple of simple cases.  First, we consider the case $(\alpha_1,\alpha_2,\alpha_3,\alpha_4)=(1,1,1,\alpha)$.  The difference between the steady-state growth rates of the synchronously and asynchronously dividing subpopulations, $\phi^*_x(1,1,1,\alpha)$ and $\phi^*_y(1,1,1,\alpha)$, respectively, are plotted in Figure \ref{fig:alpha_4}A.  If $\alpha=1$, then $\phi^*_x(1,1,1,\alpha)$ and $\phi^*_y(1,1,1,\alpha)$ are exactly equal, as already noted.  If $\alpha<1$, then $\phi^*_x(1,1,1,\alpha)<\phi^*_y(1,1,1,\alpha)$, and asynchronously dividing cells outcompete synchronously dividing cells.  But if $\alpha<1$, then a solitary phenotype would outcompete the staying-together phenotype, and there would be no formation of clusters in the first place.  If $\alpha>1$, then $\phi^*_x(1,1,1,\alpha)>\phi^*_y(1,1,1,\alpha)$, and synchronously dividing cells outcompete asynchronously dividing cells.  Thus, for $\alpha>1$, multicellularity with synchronized cell division evolves.

What is the intuition for this result?  To develop insight, we return to the simple case of $\alpha=1$.  In steady state, what fraction of the total number of cells in the synchronously and asynchronously dividing subpopulations belong to $4$-complexes?  The steady-state distributions of cluster sizes are shown in Figure \ref{fig:n=4}.  For the synchronous phenotype, exactly $1/2$ of all cells belong to $4$-complexes in steady state.  For the asynchronous phenotype, exactly $12/25$ ($<1/2$) of all cells belong to $4$-complexes in steady state.  The fraction of cells that belong to $4$-complexes---and are therefore affected by small changes in $\alpha$---is larger for the synchronous phenotype.  This suggests that, if $\alpha$ differs from $1$ by a small amount, then the corresponding effect on the population's fitness---either positive or negative---is amplified for synchronously dividing cells.

For example, if $\alpha=1+\epsilon$ with $0 < \epsilon \ll 1$, then approximately $1/2$ of all synchronously dividing cells produce new cells at an enhanced rate $1+\epsilon$, while only approximately $12/25$ of all asynchronously dividing cells produce new cells at the same enhanced rate $1+\epsilon$.  In this case, synchronization is the more successful phenotype.  If, instead, $\alpha=1-\epsilon$ with $0 < \epsilon \ll 1$, then approximately $1/2$ of all synchronously dividing cells produce new cells at a reduced rate $1-\epsilon$, while only approximately $12/25$ of all asynchronously dividing cells produce new cells at the same reduced rate $1-\epsilon$.  In this case, asynchronization is the more successful phenotype.

Our intuition further suggests that, for different values of $(\alpha_1,\alpha_2,\alpha_3,\alpha_4)$, the asynchronous phenotype can outcompete the synchronous phenotype, under conditions in which multicellularity will evolve.  Consider again the steady-state distributions of $k$-complexes for the case of $\alpha=1$, as shown in Figure \ref{fig:n=4}.  What fraction of all cells in the synchronous and asynchronous subpopulations belong to $k$-complexes of size $k \geq 2$?  In steady state, $3/4$ of all synchronously dividing cells belong to complexes with at least $2$ cells, while $19/25$ ($>3/4$) of all asynchronously dividing cells belong to complexes with at least $2$ cells.  Therefore, we anticipate that, for the fitness values $(\alpha_1,\alpha_2,\alpha_3,\alpha_4)=(1,\alpha,\alpha,\alpha)$, and for $\alpha=1+\epsilon$ with $0 < \epsilon \ll 1$, the asynchronous phenotype is more successful, and multicellularity is also evolutionarily preferred over the solitary phenotype.

The difference between the steady-state growth rates of the synchronously and asynchronously dividing subpopulations, $\phi^*_x(1,\alpha,\alpha,\alpha)$ and $\phi^*_y(1,\alpha,\alpha,\alpha)$, respectively, are plotted in Figure \ref{fig:alpha_4}B.  Our expectation is correct:  If $\alpha<1$, then the synchronous phenotype outcompetes the asynchronous phenotype, but a solitary phenotype would also outcompete a staying-together phenotype, and there would be no formation of clusters.  If $\alpha>1$, then the asynchronous phenotype outcompetes the synchronous phenotype, and evolutionary construction develops.

\subsection{$n=8$ cells}

We can also consider the case $n=8$, for which a complex contains a maximum of $8$ cells.  The dynamics of asynchronous cell division and staying together for $n=8$ are described by the following equations:
\begin{equation}
\begin{pmatrix} \dot{y}_1 \\ \dot{y}_2 \\ \dot{y}_3 \\ \dot{y}_4 \\ \dot{y}_5 \\ \dot{y}_6 \\ \dot{y}_7 \\ \dot{y}_8 \end{pmatrix} = \begin{pmatrix} -\alpha_1 & 0 & 0 & 0 & 0 & 0 & 0 & 8\alpha_8 \\ \alpha_1 & -2\alpha_2 & 0 & 0 & 0 & 0 & 0 & 0 \\ 0 & 2\alpha_2 & -3\alpha_3 & 0 & 0 & 0 & 0 & 0 \\ 0 & 0 & 3\alpha_3 & -4\alpha_4 & 0 & 0 & 0 & 0 \\ 0 & 0 & 0 & 4\alpha_4 & -5\alpha_5 & 0 & 0 & 0 \\ 0 & 0 & 0 & 0 & 5\alpha_5 & -6\alpha_6 & 0 & 0 \\ 0 & 0 & 0 & 0 & 0 & 6\alpha_6 & -7\alpha_7 & 0 \\ 0 & 0 & 0 & 0 & 0 & 0 & 7\alpha_7 & 0 \end{pmatrix} \begin{pmatrix} y_1 \\ y_2 \\ y_3 \\ y_4 \\ y_5 \\ y_6 \\ y_7 \\ y_8 \end{pmatrix} - \phi_y(\vec{\alpha};\vec{y}) \begin{pmatrix} y_1 \\ y_2 \\ y_3 \\ y_4 \\ y_5 \\ y_6 \\ y_7 \\ y_8 \end{pmatrix}.
\label{eqn:n=8_asynch}
\end{equation}
We choose $\phi_y(\vec{\alpha};\vec{y})$ such that the total number of cells equals one at all times.  We obtain
\begin{equation}
\phi_y(\vec{\alpha};\vec{y}) = 1 + \sum_{k=1}^8 ky_k(\alpha_k-1).
\label{eqn:phi_A_8}
\end{equation}
The dynamics of synchronous cell division and staying together for $n=8$ are described by the following equations:
\begin{equation}
\begin{pmatrix} \dot{x}_1 \\ \dot{x}_2 \\ \dot{x}_4 \\ \dot{x}_8 \end{pmatrix} = \begin{pmatrix} -\alpha_1 & 0 & 0 & 8\alpha_8 \\ \alpha_1 & -\alpha_2 & 0 & 0 \\ 0 & \alpha_2 & -\alpha_4 & 0 \\ 0 & 0 & \alpha_4 & 0 \end{pmatrix} \begin{pmatrix} x_1 \\ x_2 \\ x_4 \\ x_8 \end{pmatrix} - \phi_x(\vec{\alpha};\vec{x}) \begin{pmatrix} x_1 \\ x_2 \\ x_4 \\ x_8 \end{pmatrix}.
\label{eqn:n=8_synch}
\end{equation}
We choose $\phi_x(\vec{\alpha};\vec{x})$ such that the total number of cells equals one at all times.  We obtain
\begin{equation}
\phi_x(\vec{\alpha};\vec{x}) = 1 + \sum_{p=0}^3 kx_k\left(\alpha_k-1\right), \mathrm{\; where \;} k=2^p.
\label{eqn:phi_S_8}
\end{equation}
Notice that, from Equations \eqref{eqn:phi_A_8} and \eqref{eqn:phi_S_8}, if $\alpha_k=1$ for all $k$, then $\phi_y(\vec{\alpha};\vec{y})=\phi_x(\vec{\alpha};\vec{x})=\phi_y^*(\vec{\alpha})=\phi_x^*(\vec{\alpha})=1$, and there is neutrality.

\begin{figure}
\centering
\includegraphics*[width=0.9\textwidth]{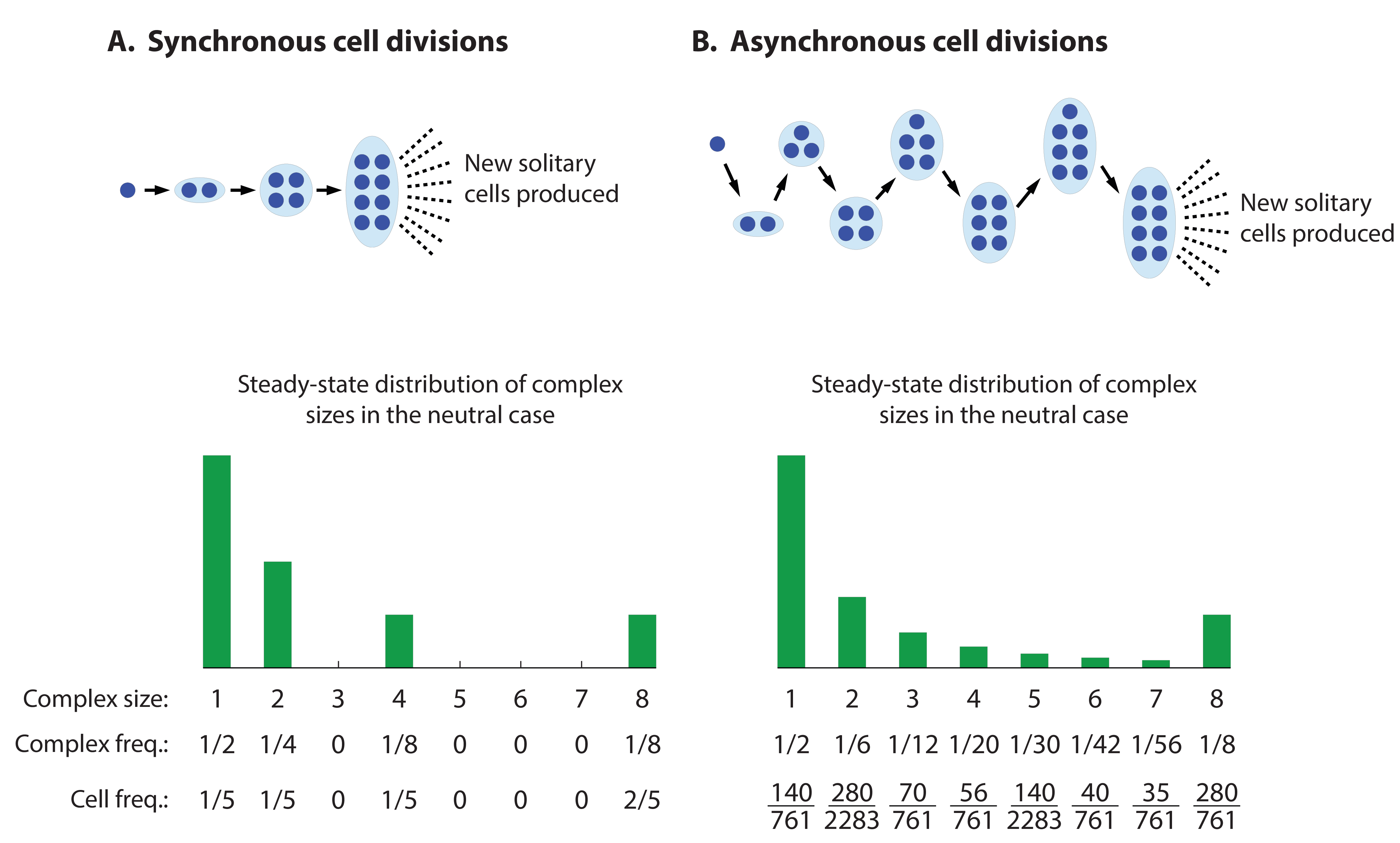}
\caption{Growth of multicellular organisms by synchronous and asynchronous cell division, when maximum size is $n=8$ cells.}
\label{fig:n=8}
\end{figure}

The processes of staying together with synchronous and asynchronous cell division for the case $n=8$ are shown schematically in Figure \ref{fig:n=8}.

\begin{figure}
\centering
\includegraphics*[width=0.7\textwidth]{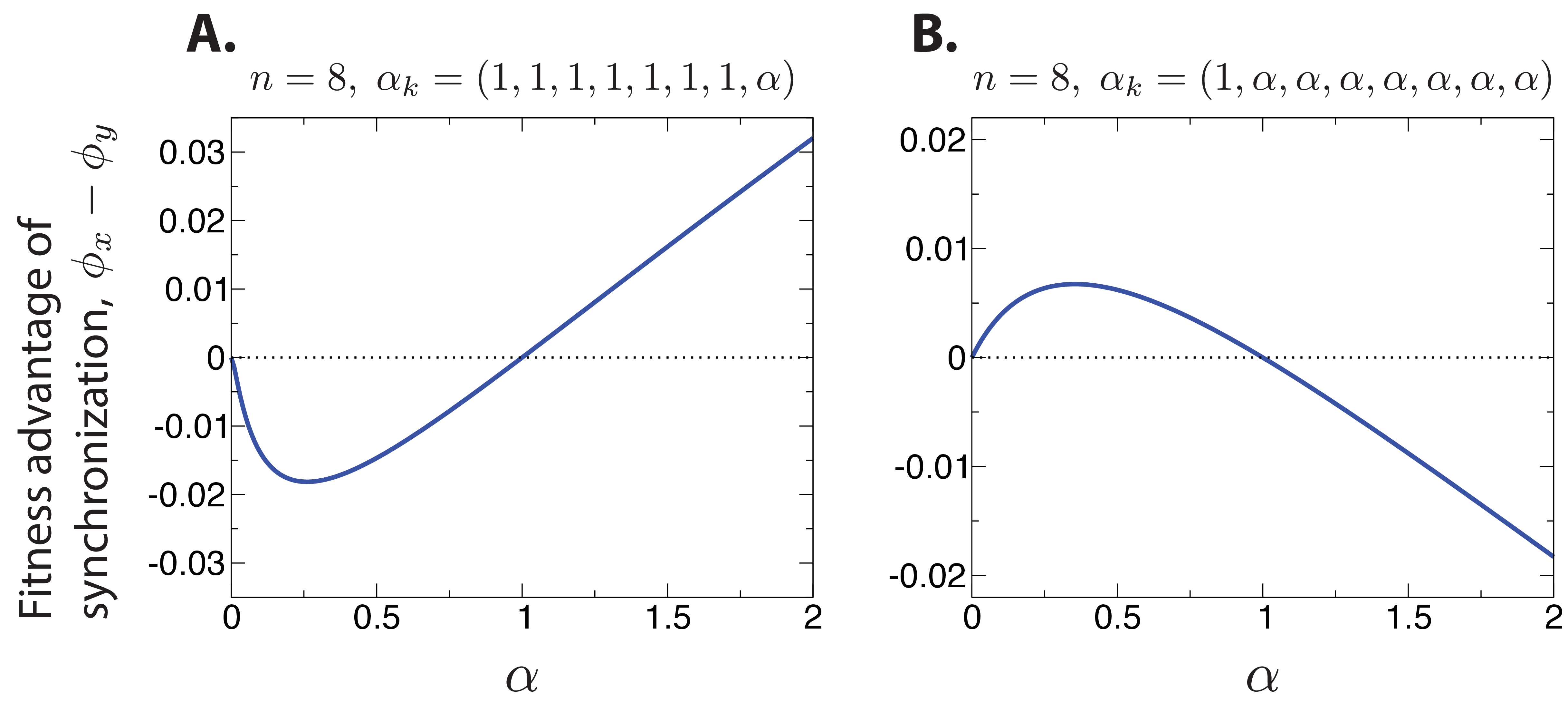}
\caption{The differences in the steady-state fitnesses of the synchronously and asynchronously dividing populations are shown for two choices of $\vec{\alpha}$ for $n=8$.  When multicellularity is selected for ($\alpha > 1$), synchronization of cell divisions is the more successful phenotype in some cases (A), and asynchronization in others (B).}
\label{fig:alpha_8}
\end{figure}

To see what this means for the evolutionary dynamics, consider the steady-state fraction of total cells in the synchronously and asynchronously dividing populations that belong to $8$-complexes if there is neutrality.  The steady-state distributions of cluster sizes are shown in Figure \ref{fig:n=8}.  For the synchronous case, that fraction is $2/5$, while for the asynchronous case, that fraction is $280/761\approx 0.367$ ($<2/5$).  This suggests that, for the case $\vec{\alpha}=(1,1,1,1,1,1,1,\alpha)$, and for $\alpha=1+\epsilon$ with $0 < \epsilon \ll 1$, multicellularity is selected and the synchronous phenotype is more successful.  If instead $\alpha=1-\epsilon$ with $0 < \epsilon \ll 1$, then the asynchronous phenotype would be more successful, but multicellularity is not selected.  As shown in Figure \ref{fig:alpha_8}A, our expectation is correct.

Consider also the steady-state fraction of total cells in the synchronously and asynchronously dividing populations that belong to complexes containing at least $2$ cells if there is neutrality.  For the synchronous case, that fraction is $4/5$, while for the asynchronous case, that fraction is $621/761 \approx 0.816$ ($>4/5$).  This suggests that, for the case $\vec{\alpha}=(1,\alpha,\alpha,\alpha,\alpha,\alpha,\alpha,\alpha)$, and for $\alpha=1+\epsilon$ with $0 < \epsilon \ll 1$, multicellularity is selected and the asynchronous phenotype is more successful.  If instead $\alpha=1-\epsilon$ with $0 < \epsilon \ll 1$, then the synchronous phenotype would be more successful, but multicellularity is not selected.  As shown in Figure \ref{fig:alpha_8}B, our expectation is again correct.

\subsection{Any number of cells}

We have considered the cases $n=4$ and $n=8$, but more generally, we can describe the evolutionary dynamics for any value of the maximum complex size, $n$.  The dynamics of asynchronous cell division and staying together are described by the following equations:
\begin{equation*}
\begin{aligned}
\dot{y}_1 &= - \alpha_1y_1 + n\alpha_ny_n - \phi_y(\vec{\alpha};\vec{y}) y_1; \\
\dot{y}_k &= (k-1)\alpha_{k-1}y_{k-1} - k\alpha_ky_k - \phi_y(\vec{\alpha};\vec{y}) y_k \quad \forall \quad 1 < k < n; \\
\dot{y}_n &= (n-1)\alpha_{n-1}y_{n-1} - \phi_y(\vec{\alpha};\vec{y}) y_n. \\
\end{aligned}
\end{equation*}
We choose $\phi_y(\vec{\alpha};\vec{y})$ such that the total number of cells equals one at all times.  We obtain
\begin{equation}
\phi_y(\vec{\alpha};\vec{y}) = 1 + \sum_{k=1}^n ky_k(\alpha_k-1).
\label{eqn:phi_A_n}
\end{equation}
The dynamics of synchronous cell division and staying together are described by the following equations:
\begin{equation*}
\begin{aligned}
\dot{x}_1 &= - \alpha_1x_1 + n\alpha_nx_n - \phi_x(\vec{\alpha};\vec{x}) x_1; \\
\dot{x}_k &= \alpha_{k/2}x_{k/2} - \alpha_kx_k - \phi_x(\vec{\alpha};\vec{x}) x_k \quad \forall \quad k=2^p, \quad 0 < p < \log_2 n; \\
\dot{x}_n &= \alpha_{n/2}x_{n/2} - \phi_x(\vec{\alpha};\vec{x}) x_n. \\
\end{aligned}
\end{equation*}
We choose $\phi_x(\vec{\alpha};\vec{x})$ such that the total number of cells equals one at all times.  We obtain
\begin{equation}
\phi_x(\vec{\alpha};\vec{x}) = 1 + \sum_{p=0}^{\log_2 n} kx_k\left(\alpha_k-1\right), \mathrm{\; where \;} k=2^p.
\label{eqn:phi_S_n}
\end{equation}
From Equations \eqref{eqn:phi_A_n} and \eqref{eqn:phi_S_n}, it follows that, if $\alpha_k=1$ for all $k$, then $\phi_y(\vec{\alpha};\vec{y})=\phi_x(\vec{\alpha};\vec{x})=\phi_y^*(\vec{\alpha})=\phi_x^*(\vec{\alpha})=1$, and there is neutrality.  But the same intuition that applies for the cases $n=4$ and $n=8$ also applies for larger values of $n$:  If $\alpha_k \neq 1$ for some value of $k$, then, in general, natural selection will act differently on synchronous and asynchronous phenotypes.

\subsection{Different fitnesses for cells of asynchronous and synchronous phenotypes}

The consideration of the same fitness, $\alpha_k$, of cells in a $k$-complex between the asynchronous and synchronous phenotypes thus reveals that natural selection acts differently on the two modes of cellular division.  More generally, one can consider that cells in $k$-complexes that divide asynchronously versus synchronously do not necessarily have the same fitness.

Denote by $\beta_k$ the fitness of cells in an asynchronously dividing $k$-complex.  We have the following dynamics:
\begin{equation*}
\begin{aligned}
\dot{Y}_1 &= - \beta_1Y_1 + n\beta_nY_n - \Phi_Y(\vec{\beta};\vec{Y}) Y_1; \\
\dot{Y}_k &= (k-1)\beta_{k-1}Y_{k-1} - k\beta_kY_k - \Phi_Y(\vec{\beta};\vec{Y}) Y_k \quad \forall \quad 1 < k < n; \\
\dot{Y}_n &= (n-1)\beta_{n-1}Y_{n-1} - \Phi_Y(\vec{\beta};\vec{Y}) Y_n. \\
\end{aligned}
\end{equation*}
We choose $\Phi_Y(\vec{\beta};\vec{Y})$ such that the total number of cells equals one at all times.  We obtain
\begin{equation*}
\Phi_Y(\vec{\beta};\vec{Y}) = 1 + \sum_{k=1}^n kY_k(\beta_k-1).
\end{equation*}
Denote by $\gamma_k$ the fitness of cells in a synchronously dividing $k$-complex.  We have the following dynamics:
\begin{equation*}
\begin{aligned}
\dot{X}_1 &= - \gamma_1X_1 + n\gamma_nX_n - \Phi_X(\vec{\gamma};\vec{X}) X_1; \\
\dot{X}_k &= \gamma_{k/2}X_{k/2} - \gamma_kX_k - \Phi_X(\vec{\gamma};\vec{X}) X_k \quad \forall \quad k=2^p, \quad 0 < p < \log_2 n; \\
\dot{X}_n &= \gamma_{n/2}X_{n/2} - \Phi_X(\vec{\gamma};\vec{X}) X_n. \\
\end{aligned}
\end{equation*}
We choose $\Phi_X(\vec{\gamma};\vec{X})$ such that the total number of cells equals one at all times.  We obtain
\begin{equation*}
\Phi_X(\vec{\gamma};\vec{X}) = 1 + \sum_{p=0}^{\log_2 n} kX_k\left(\gamma_k-1\right), \mathrm{\; where \;} k=2^p.
\end{equation*}
Here, in general, $\beta_k \neq \gamma_k$.  What is the condition for the synchronous phenotype to be favored over the asynchronous phenotype, or vice versa?

For the asynchronously dividing phenotype, setting $\dot{Y}_k=0$ for all $k$, we have the following homogeneous linear system:
\begin{equation*}
0 = \sum_{j=1}^n B_{ij}(\vec{\beta},\Phi_Y^*(\vec{\beta})) Y_j^* \quad \forall \quad 1 \leq i \leq n.
\end{equation*}
For the synchronously dividing phenotype, setting $\dot{X}_k=0$ for all $k$, we have the following homogeneous linear system:
\begin{equation*}
0 = \sum_{j=1}^{1 + \log_2 n} C_{ij}(\vec{\gamma},\Phi_X^*(\vec{\gamma})) X_k^* \quad \mathrm{where} \quad k = 2^{j-1} \quad \forall \quad 1 \leq i \leq 1 + \log_2 n.
\end{equation*}

For the asynchronously dividing phenotype, using a cofactor expansion, we can solve implicitly for $\Phi_Y^*$:
\begin{equation}
\det(B_{ij}(\vec{\beta},\Phi_Y^*(\vec{\beta}))) = \Phi_Y^* \prod_{i=1}^{n-1} \left(\Phi_Y^* + i\beta_i\right) - \prod_{i=1}^n i\beta_i = 0.
\label{eqn:solution_Phi_Y}
\end{equation}
For the synchronously dividing phenotype, using a cofactor expansion, we can solve implicitly for $\Phi_X^*$:
\begin{equation}
\det(C_{ij}(\vec{\gamma},\Phi_X^*(\vec{\gamma}))) = \Phi_X^* \prod_{i=1}^{\log_2 n} \left(\Phi_X^* + \gamma_k\right) - n\gamma_n \prod_{i=1}^{\log_2 n} \gamma_k = 0 \quad \mathrm{where} \quad k = 2^{i-1}.
\label{eqn:solution_Phi_X}
\end{equation}

We are interested in the largest real values of $\Phi_Y^*$ and $\Phi_X^*$ that satisfy Equations \eqref{eqn:solution_Phi_Y} and \eqref{eqn:solution_Phi_X}, respectively.  If $\Phi_Y^*>\Phi_X^*$, then the asynchronous phenotype outcompetes the synchronous phenotype.  If $\Phi_Y^*<\Phi_X^*$, then the synchronous phenotype outcompetes the asynchronous phenotype.  If $\Phi_Y^*=\Phi_X^*$, then the asynchronous and synchronous phenotypes coexist.

\subsection{Weak selection}

We can further consider the simplified case in which $\beta_k = 1 + \epsilon (\delta\beta_k)$ and $\gamma_k = 1 + \epsilon (\delta\gamma_k)$, where $0 < \epsilon \ll 1$.  Accordingly, we have $\Phi_Y^* = 1 + \epsilon (\delta\Phi_Y^*)$ and $\Phi_X^* = 1 + \epsilon (\delta\Phi_X^*)$.  With these substitutions, Equation \eqref{eqn:solution_Phi_Y} becomes
\begin{equation*}
(\delta\Phi_Y^*) \sum_{i=1}^n \frac{1}{i} + \sum_{i=1}^{n-1} \frac{i}{i+1} (\delta\beta_i) - \sum_{i=1}^n (\delta\beta_i) = 0.
\end{equation*}
We can solve explicitly for $\delta\Phi_Y^*$:
\begin{equation}
\delta\Phi_Y^* = \frac{\sum_{i=1}^{n-1} \frac{1}{i+1} (\delta\beta_i) + (\delta\beta_n)}{\sum_{i=1}^n \frac{1}{i}}.
\label{eqn:delta_Phi_Y_weak}
\end{equation}
Also, Equation \eqref{eqn:solution_Phi_X} becomes
\begin{equation*}
(\delta\Phi_X^*) \left( 1 + \frac{\log_2 n}{2} \right) + \frac{1}{2} \sum_{i=1}^{\log_2 n} \delta\gamma_k - \sum_{i=1}^{1 + \log_2 n} \delta\gamma_k = 0 \quad \mathrm{where} \quad k = 2^{i-1}.
\end{equation*}
We can solve explicitly for $\delta\Phi_X^*$:
\begin{equation}
\delta\Phi_X^* = \frac{\sum_{i=1}^{\log_2 n} (\delta\gamma_k) + 2(\delta\gamma_n)}{2 + \log_2 n} \quad \mathrm{where} \quad k = 2^{i-1}.
\label{eqn:delta_Phi_X_weak}
\end{equation}
If $\delta\Phi_Y^*>\delta\Phi_X^*$, then the asynchronous phenotype outcompetes the synchronous phenotype.  If $\delta\Phi_Y^*<\delta\Phi_X^*$, then the synchronous phenotype outcompetes the asynchronous phenotype.  If $\delta\Phi_Y^*=\delta\Phi_X^*$, then the asynchronous and synchronous phenotypes coexist.

\subsubsection{Particular case}

As an example, consider $\vec{\beta}=\vec{\gamma}$ as a monotonically increasing function of the number of cells, $k$, in a $k$-complex.  For $0 < \epsilon \ll 1$, one possibility is:
\begin{equation}
\gamma_k = \beta_k = \alpha_k = 1 + \epsilon \left( \frac{k-1}{n-1} \right) ^c.
\label{eqn:alpha_particular}
\end{equation}
For the particular choice of $\vec{\alpha}$ given by Equation \eqref{eqn:alpha_particular}, and for $n=4$, the parameter $c$ can be thought of as interpolating between the cases $(\alpha_1,\alpha_2,\alpha_3,\alpha_4)=(1,\epsilon,\epsilon,\epsilon)$ in the limit $c \rightarrow 0$ and $(\alpha_1,\alpha_2,\alpha_3,\alpha_4)=(1,1,1,\epsilon)$ in the limit $c \rightarrow \infty$, which facilitates comparison with the results described above.

For this example, we restrict our attention to positive values of $\epsilon$, which ensures that multicellularity evolves in the first place.  For sufficiently small values of $c$, $\alpha_k$ is a concave function of $k$, and one might expect evolution of asynchronous cell division, while for sufficiently large values of $c$, $\alpha_k$ is a convex function of $k$, and one might expect evolution of synchronous cell division.  Is there a critical value of $c$ above which the synchronous phenotype becomes preferred over the asynchronous phenotype?

Using Equation \eqref{eqn:delta_Phi_Y_weak}, we have
\begin{equation*}
\delta\Phi_Y^* = \frac{\frac{1}{3} \left(\frac{1}{3}\right)^c + \frac{1}{4} \left(\frac{2}{3}\right)^c + 1}{1 + \frac{1}{2} + \frac{1}{3} + \frac{1}{4}}.
\end{equation*}
Using Equation \eqref{eqn:delta_Phi_X_weak}, we have
\begin{equation*}
\delta\Phi_X^* = \frac{\left(\frac{1}{3}\right)^c + 2}{4}.
\end{equation*}
Setting $\delta\Phi_Y^*=\delta\Phi_X^*$ and simplifying, we can solve implicitly for the critical value of $c$:
\begin{equation}
4\left(\frac{2}{3}\right)^c = 3\left(\frac{1}{3}\right)^c + \frac{2}{3}
\label{eqn:alpha_particular_n=4}
\end{equation}

The numerical solution of Equation \eqref{eqn:alpha_particular_n=4} is $c \approx 4.32$.  If $c \lesssim 4.32$, then the asynchronous phenotype wins over the synchronous phenotype.  If $c \gtrsim 4.32$, then the synchronous phenotype wins over the asynchronous phenotype.  Thus, for $n=4$, $\alpha_k$ must increase very sharply with $k$ around $k=4$ for synchronization of cell division times to evolve.

\section{Discussion}

We have studied a simple model in which there compete multicellular organisms that grow either by synchronous or asynchronous cell division. We have shown that, under certain conditions, selection favors the synchronous phenotype. The basic intuition is that an organism growing by synchronous cell division bypasses certain sizes in terms of cell number---specifically, those that are not a power of 2---and if these bypassed cell sizes are relatively unproductive, then synchronous cell division can be favored.

More specifically, the effect of synchrony is to alter the frequencies of the various organism sizes in the population's steady-state distribution, which, in general, alters the population's growth rate in steady state.  If the fitness of a $k$-complex, $\alpha_k$, is equal to $1$ for all $k$, then there is neutrality between the synchronous and asynchronous phenotypes.  But the case $\alpha_k=1$ for all $k$ is nongeneric.  For example, for $n=4$, we have demonstrated that if $\vec{\alpha}=(1,1,1,1+\epsilon)$ with $0 < \epsilon \ll 1$, then synchronization is the more successful phenotype.  The intuition is that, in steady state, $1/2$ of all synchronously dividing cells belong to $4$-complexes, while only $12/25$ of all asynchronously dividing cells belong to $4$-complexes.  So a greater fraction of cells of the synchronous phenotype exhibit an enhanced fitness $1+\epsilon$ compared with the asynchronous phenotype, and synchronization in this setup is favored.  An intriguing possibility is that solitary cells are less fit than clusters of any size greater than one, i.e., $\vec{\alpha}=(1-\epsilon,1,1,1)$ for $0 < \epsilon \ll 1$.  The same intuition applies:  In steady state, $1/4$ of all synchronously dividing cells are solitary, while only $6/25$ of all asynchronously dividing cells are solitary.  So a greater fraction of cells of the synchronous phenotype exhibit a reduced fitness $1-\epsilon$ compared with the asynchronous phenotype, and asynchronization in this setup is favored.

Therefore, importantly, the evolutionary success of synchrony over asynchrony does not require unrealistic situations where, for example, organisms of sizes 2, 4, and 8 cells are very efficient at producing new cells but organisms of sizes 3, 5, 6, and 7 cells are inefficient. The relative efficiency of synchrony can in fact be ensured under sensible specifications of the efficiency of different organism sizes, for example, the monotonically increasing convex specification in Eq.~\eqref{eqn:alpha_particular}.

Many other realistic specifications can be imagined. For example, if the benefit to multicellularity is that it reduces the diffusion of heat or some chemical out of the complex by increasing the ratio of volume to surface area, then the fitness advantage of multicellularity would seem to grow with complex size $k$ according to this ratio, viz.~$\alpha_k \sim k^{1/3}$. For the case $\alpha_k = k^{1/3}$, for both $n=4$ and $n=8$, numerical solutions of Equations \eqref{eqn:solution_Phi_Y} and \eqref{eqn:solution_Phi_X} reveal that the asynchronous phenotype is more successful than the synchronous phenotype.

A possible consideration is that synchronous cell division, whatever its benefits in principle, is in practice difficult to achieve. But many instances of synchronized cell division can be found in nature. Early development in animals is characterized by rapid synchronous cell divisions [for example, the first 11 or 12 cell divisions of \textit{Xenopus} embryos are synchronous \citep{Satoh_1977,Newport_1982}]. In \textit{Xenopus}, these synchronous early cell divisions are the result of clock-like cycles of activation and inactivation of the kinase protein Cdc2, driven by an autonomous negative-feedback loop between the cyclin synthesis (which activates Cdc2) and the anaphase-promoting complex \citep{Murray_1989,Pomerening_2005}. This indicates that synchronous cell divisions in a multicellular organism are quite possible, and suggest a mechanism by which they may be achieved: biochemical negative-feedback loops. (Later cell divisions of an animal embryo are not synchronous, as it undergoes gastrulation and tissue differentiation, but these are stages that would not characterize the simple (undifferentiated) multicellular organisms we have modeled.)

Moreover, synchronously dividing cultures of bacteria and other microbes have been produced experimentally by first enforcing a stationary phase in which cells undergo no new rounds of cell division, and then suddenly inducing cell divisions by enriching the environment of the culture. Several subsequent rounds of cell division are then approximately synchronous across the culture \citep{Cutler_1966}. This suggests that basing cell division on an approximately periodic intracellular cue could lead to synchrony in the early cell divisions of a multicellular organism, as required by our model. For example, if DNA replication is continuous, then it can lead to synchronized cell divisions if it proceeds at an approximately constant speed across cells. Alternatively, if cell volumes grow at an approximately constant speed across cells, then the use of a certain threshold cell volume as the signal to divide would lead to approximate synchrony in cell divisions. Selection in favor of synchronous cell divisions, the conditions for which we have studied, could then lead to reliance on as precisely periodic a cue as possible, or even the entrainment of some cue onto a periodic cycle. 

\section*{Competing interests}

We have no competing interests.

\section*{Authors' contributions}

All authors performed research for this work.

\section*{Funding}

This work was supported by the Bill and Melinda Gates Foundation [OPP1148627], Office of Naval Research grant N00014-16-1-2914, National Cancer Institute grant CA179991 and by the John Templeton Foundation. The Program for Evolutionary Dynamics is supported in part by a gift from B Wu and Eric Larson.

\bibliographystyle{plainnat}
\bibliography{synch}

\end{document}